\documentclass[]{fairmeta}

\makeatletter
\let\fairmeta@usepackage\usepackage
\renewcommand{\usepackage}[2][]{%
  \ifstrequal{#2}{subfigure}{}{%
    \ifstrequal{#2}{tcolorbox}{}{%
      \fairmeta@usepackage[#1]{#2}}}}
\usepackage{microtype}
\usepackage{amsmath,amsfonts,bm,bbm}
\usepackage{amsthm}
\usepackage{graphicx}
\usepackage{caption}
\usepackage{color}
\usepackage{subfigure}
\usepackage{comment}
\usepackage[style=american]{csquotes}
\MakeOuterQuote{"} % Render coauthor-friendly "text" as typographic quotes.
\usepackage{enumitem}
\usepackage{mathtools}
\usepackage{thmtools, thm-restate}
\usepackage[ruled,linesnumbered]{algorithm2e}  
\usepackage{algorithmic}
\usepackage{verbatim}
\usepackage{cuted}

\usepackage{hyperref}
\usepackage{cleveref}
\crefname{equation}{Eq.}{Eqs.}
\Crefname{equation}{Eq.}{Eqs.}
\usepackage{xcolor}
\usepackage{natbib}
\usepackage{listings}
\usepackage{xparse}
\usepackage{blindtext}
\usepackage{lipsum}
\usepackage{caption}
\usepackage{subcaption}
\usepackage{enumitem}
\usepackage{makecell}
\usepackage{anyfontsize}
\usepackage{soul}

\newcommand{\stopgrad}{\mathop{\text{\normalfont\texttt{stop-grad}}}}
\newcommand{\innp}[1]{\left\langle #1 \right\rangle}

\theoremstyle{plain} \numberwithin{equation}{section}
\newtheorem{theorem}{Theorem}[section]
\numberwithin{theorem}{section}

\newtheorem{fact}[theorem]{Fact}
\theoremstyle{definition}

\usepackage[breakable, skins]{tcolorbox}
\definecolor{greyC}{RGB}{180,180,180}
\definecolor{greyL}{RGB}{235,235,235}

\def\1{\mathbbm{1}}

\def\vone{{\bm{1}}}

\def\vc{{\bm{c}}}

\def\ve{{\mathbf{e}}}
\def\vf{{\bm{f}}}
\def\vg{{\mathbf{g}}}

\def\vi{{\bm{i}}}
\def\vj{{\bm{j}}}

\def\vp{{\bm{p}}}
\def\vq{{\bm{q}}}
\def\vr{{\bm{r}}}

\def\vu{{\mathbf{u}}}
\def\vv{{\mathbf{v}}}
\def\vx{{\mathbf{x}}}

\def\vz{{\bm{z}}}

\def\mC{{\bm{C}}}

\def\mU{{\bm{U}}}
\def\mV{{\bm{V}}}

\DeclareMathAlphabet{\mathsfit}{\encodingdefault}{\sfdefault}{m}{sl}
\SetMathAlphabet{\mathsfit}{bold}{\encodingdefault}{\sfdefault}{bx}{n}

\def\gB{{\mathcal{B}}}

\def\gL{{\mathcal{L}}}

\def\gS{{\mathcal{S}}}

\newcommand{\R}{\mathbb{R}}

\newcommand{\softmax}{\mathrm{softmax}}

\newcommand{\KL}{\mathbb{D}_{\mathrm{KL}}}

\DeclareMathOperator*{\argmax}{arg\,max}

\let\usepackage\fairmeta@usepackage
\makeatother

\AtBeginDocument{\let\cite\citep}

\usepackage{float}
\usepackage{adjustbox}
\graphicspath{{../}}

\AtBeginEnvironment{table}{%
  \small
  \let\resizebox\arxivmaxwidth
}

\let\arxivoriginaltabular\tabular
\let\arxivoriginalendtabular\endtabular
\newcommand{\arxivwidetabular}[1]{%
  \setlength{\tabcolsep}{2pt}%
  \arxivoriginaltabular{@{}%
    >{\raggedright\arraybackslash}p{0.125\textwidth}%
    >{\raggedright\arraybackslash}p{0.245\textwidth}%
    *{5}{>{\centering\arraybackslash}p{0.107\textwidth}}@{}}%
}

\AtBeginEnvironment{table*}{%
  \small
  \let\resizebox\arxivnoresize
  \let\tabular\arxivwidetabular
  \let\endtabular\arxivoriginalendtabular
}

\newcommand{\oneshot}{OneShot}

\AtBeginDocument{%
  }

\title{OneShot: Index-in-Ranking with Neural Scoring \\ for Large-Scale Retrieval}

\author[*]{Ziwei Li}
\author[*]{Shuyao Li}
\author[*]{Xufeng Cai}
\author[*]{Xue Zou}
\author[*]{Yiming Ma}
\author[\dagger]{Huiting Lu}
\author[\dagger]{Wujie Yan}
\author[\dagger]{Zhichen Zhao}
\author[*]{Yang Lu}
\author[*]{Zhe Wang}
\author[*]{Rui Luo}
\author[*]{Zhengyu Su}
\author[*]{Dan Zhang}
\author[*]{Yimin Tan}
\author[*]{Ji Liu}

\affiliation{Meta Platforms, Inc. $^{*}$USA, $^{\dagger}$Singapore}

\metadata[Email]{\texttt{\{ziweili, shuyaoli, xufengcai, xuez, yma007, huiting, willyan, waterzhao, yanglv, zhewangjoe, luorui, zhengyu, danzhang, kevineetan, madisonliu\}@meta.com}}

\abstract{In modern recommendation systems, retrieval serves as a primary stage responsible for filtering billions of candidate items down to thousands prior to refined ranking.
To make this massive search effective and efficient, the system relies on ranking accuracy and indexing efficiency.
However, these two objectives are traditionally misaligned: while the former optimizes for the alignment between ranking predictions and user behavior, the latter optimizes for a structural grouping of item representations which enables fast search among billions of candidates.
Thus, despite extensive efforts to scale up interaction modeling for retrieval, they remain fundamentally limited by the structural misalignment between the ranking objectives and the proximity-learned index.
In this work, we address this long-standing dichotomy by proposing a new holistic retrieval framework, \oneshot.
It is an end-to-end, in-model index learning framework that natively aligns index learning with ranking objectives.
Using this joint learning as a structural foundation, \oneshot\ pushes the boundaries of retrieval expressiveness by scaling interaction modeling with neural scoring beyond the persistent dot-product bottleneck.
\oneshot\ is fully deployed in Instagram's industrial short-video recommendation system, driving significant wins in user daily sessions, engagement, and time-spent.
Additionally, \oneshot\ achieves a $20\%$ recall gain at the operational ranking volume and a 10x efficiency improvement at an equivalent recall level.
}

\begin{document}

\maketitle

\section{Introduction}\label{sec:introduction}
% Overview of recommendation pipeline
Modern recommendation systems operate at an extremely large scale, typically relying on a cascaded architecture to predict user interactions with items~\citep{covington_2016}.
The core pipeline primarily consists of two stages: retrieval and ranking.
Given a user's context and engagement history, the retrieval stage efficiently filters the candidate corpus from billions of items down to a few thousand highly relevant and manageable candidates.
Then, a more computationally intensive ranking stage evaluates these candidates, narrowing them down to dozens of items presented to the user.

% Put this table at bottom for arxiv format
\begin{table}[b]
\centering
\caption{Architectural comparison of large-scale retrieval indexing methods. \oneshot\ uniquely achieves both rank-index consistency and unconstrained interaction scale-up.}
\label{tab:method_comparison}
\begin{tabular}{lcc}
\toprule
 & \textbf{Rank-Index} & \textbf{Interaction} \\
\textbf{Method} & \textbf{consistency} & \textbf{scale-up} \\
\midrule
$k$-means ANN \citep{johnson_2017} & \texttimes & \texttimes \\
HNSW \citep{malkov_2018} / NANN \citep{chen_nann_2022} & \texttimes & \checkmark \\ %$^*$
Streaming VQ \citep{bin_2025} & \checkmark & \texttimes \\
\midrule
\oneshot & \textbf{\checkmark} & \textbf{\checkmark} \\
\bottomrule
\end{tabular}
\end{table}
\begin{figure*}[h]
    \centering
    \includegraphics[width=1.0\textwidth]{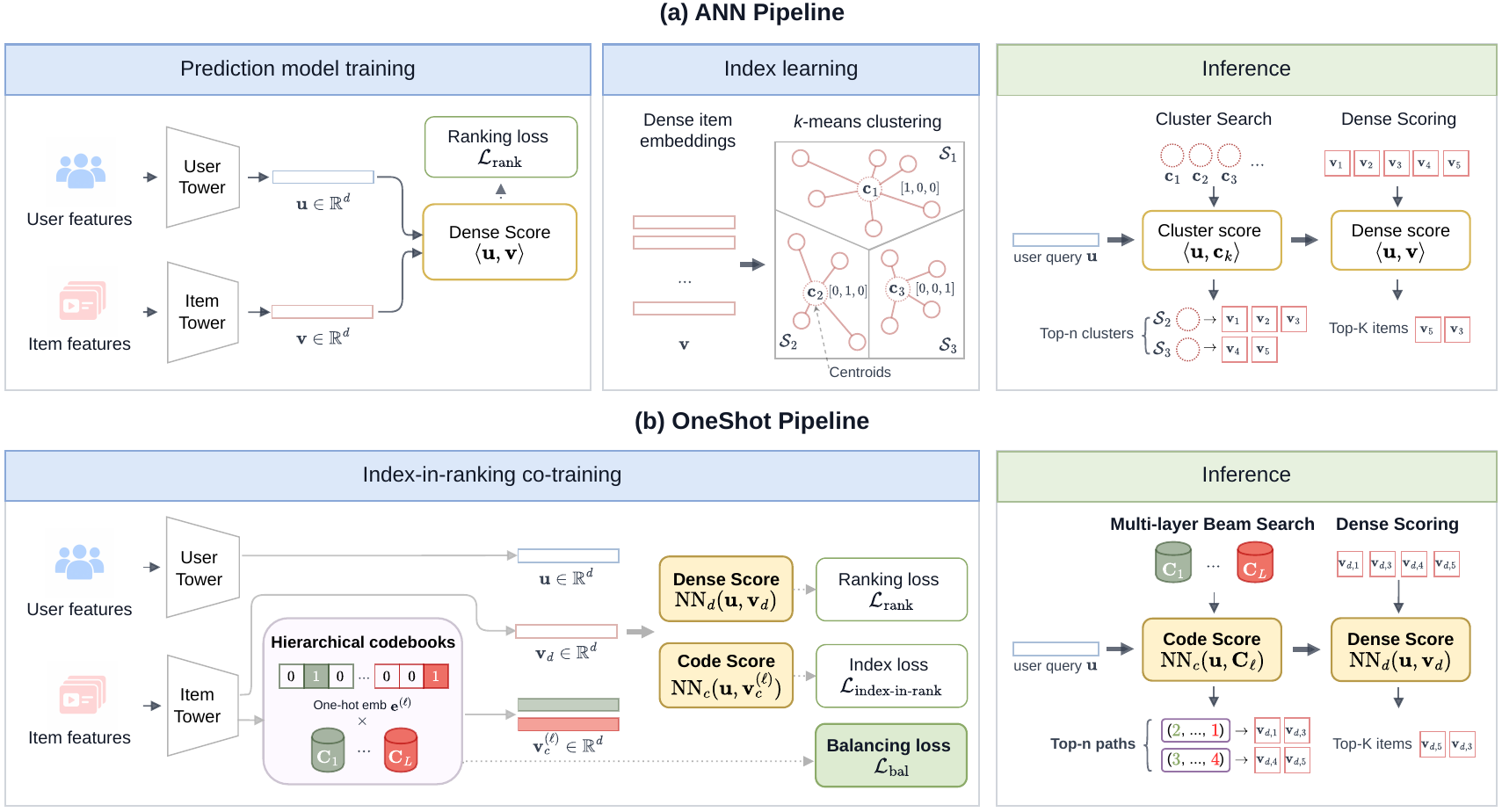}
    \caption{(a) $k$-means ANN as an example of an existing retrieval pipeline; graph-based models are similar except that their index training is hierarchical. (b) \oneshot\ multi-layer pipeline. In both panels, $\vu$ denotes user embeddings, $\vv$ and $\vv_d$ denote dense item embeddings, $\vv_c^{(\ell)}$ denotes item code embeddings, and $\mathrm{NN}_d$ and $\mathrm{NN}_c$ are arbitrary trainable neural networks. Common overheads such as item embedding caching and model packing are omitted for simplicity.}
    \label{fig:system_comparison}
\end{figure*}

To make billion-scale search tractable in the upstream retrieval stage, the system requires an index learning process that groups item representations for fast traversal, e.g., into clusters~\citep{johnson_2017, guo_2019} or spatial graphs~\citep{zhu_2018}.
Figure~\ref{fig:system_comparison}(a) shows a traditional retrieval train-serving paradigm.
In model training, the model learns to predict user-item interactions according to the ranking loss.
Then, a separate index learning process applies approximate nearest neighbor (ANN) methods such as $k$-means~\citep{johnson_2017} and HNSW~\citep{malkov_2018} to index the candidate corpus based on the spatial proximity of item embeddings.
During serving, retrieval first identifies the most relevant clusters by searching over their centroid embeddings, then evaluates only the items assigned to those clusters using their item embeddings.
Thus, retrieval fundamentally optimizes for two objectives: user-item ranking accuracy and item-only indexing efficiency.
However, these two objectives are inherently \emph{misaligned}: the ranking objective aligns model predictions with user behavior, whereas the indexing objective groups items by embedding-space proximity to maintain a balanced and representative index.

This misalignment further introduces a structural tension when scaling retrieval with more expressive interaction modeling.
Since the decoupled indexing is blind to ranking objectives, the proximity-learned centroid embeddings can only represent their clusters effectively if user-item interactions are restricted to simple metrics like dot product.
This largely constrains the retrieval framework to a late-fusion architecture with simple dot-product crossing, preventing scale-up efforts to improve retrieval expressiveness.
Consequently, efforts to scale the complexity of interaction modeling remain largely confined to downstream ranking stages~\citep{wang_2021_dcn,chang_2023_twin}.
Notably, a recent retrieval method, NANN~\citep{chen_nann_2022}, aims to scale interaction modeling by integrating a neural scoring function during graph index traversal.
However, it remains constrained by this fundamental misalignment because its neural scoring landscape on item embeddings is required to be "flat" to preserve index correctness.

Most recently, Streaming VQ~\citep{bin_2025} attempts to address this misalignment by proposing in-model indexing that attaches a vector quantization (VQ) module to the item tower, mapping continuous item embeddings to a finite, learnable codebook.
However, it uses non-differentiable codebook updates based on an exponential moving average (EMA) of item-tower outputs, so the index is not learned end-to-end (E2E) from the ranking objective.
Furthermore, its interaction modeling remains confined to a linear dot-product.
In fact,~\citet{bin_2025} explicitly reports that their indexing approach is inherently linear, preventing further scaling of interaction complexity.

\paragraph{\bf Our contributions.}
In this work, we introduce a new index-in-ranking paradigm, dubbed \oneshot, to address this long-standing index-ranking misalignment and scaling bottleneck in retrieval.
\oneshot\ fundamentally re-architects the retrieval stage through E2E in-model indexing with native neural scoring scale-up (see comparison with state-of-the-art retrievers in Table~\ref{tab:method_comparison}).
As illustrated in Figure~\ref{fig:system_comparison}(b), our pipeline unifies ranking and index learning into a single co-training stage, where index learning is directly aligned with ranking objectives.
This joint learning enables the scale-up of interaction modeling with neural scoring, resolving the historical expressiveness limit in retrieval models.
We further highlight our main technical contributions as follows:
\begin{itemize}[wide=0pt, leftmargin=\parindent]
    \item {\bf In-model hierarchical indexing. } We propose an end-to-end trainable, hierarchical in-model indexing framework that eliminates the misalignment between indexing and ranking objectives.
    This indexing scheme trains multi-layer one-hot encodings under ranking objectives alongside a boosting mechanism for refined indexing accuracy.
    We further complement this multi-layer index with efficient serving strategies based on beam search.
    \item {\bf Interaction modeling scale-up. } We incorporate neural scoring over all user-item interactions, building upon the elimination of the index-ranking misalignment.
    % -- historically restricted to simple dot products --
    This increases the expressiveness of retrieval interaction modeling and unlocks scaling opportunities previously confined to downstream ranking stages.
    \item {\bf Global index balancing. } We introduce a theoretically grounded global index-balancing method to prevent index collapse and ensure indexing efficiency.
    Derived from stochastic compositional optimization theory, our method formulates a tractable surrogate for enforcing a uniform cluster-size distribution over the global candidate corpus in batched training.
    This method is validated by significantly improved recall and cluster-size statistics compared to existing VQ methods.
\end{itemize}
\oneshot\ has been extensively validated in the Instagram short-video recommendation system, achieving a 20\% offline recall improvement over the production baseline.
Further, \oneshot\ has been fully deployed to global traffic, driving significant online gains in user engagement ($+0.04\%$ sessions, $+0.14\%$ watch time) and a $+61.6\%$ increase in the retrieval source contribution rate.
% To the best of our knowledge, this is the first fully differentiable in-model indexing framework E2E trained with ranking objective and neural scoring that is deployed in an industrial online-learning environment.
To the best of our knowledge, our work represents the first industrial design and deployment of an in-model indexing framework that is end-to-end trainable under ranking objectives alongside interaction scale-up via neural scoring.

\section{Related Work}
Traditional retrieval paradigms embed users and items into a shared latent space and rely on a decoupled indexing process for ANN search.
Notable ANN techniques include product quantization~\citep{ge2013optimized,jegou_2011}, tree-based algorithms~\citep{houle2014rank,muja2014scalable}, locality-sensitive hashing~\citep{shrivastava2014asymmetric,spring_2017}, and graph-based methods like NSW~\citep{malkov2014approximate} and HNSW~\citep{malkov_2018}.
Within decoupled indexing frameworks, there are also works utilizing multi-index binary hashing~\citep{kang2019candidate,medini2019extreme} instead of ANN search.
Several works extend embedding-based ANN search to learned similarities over user-item pairs, including tree-based methods such as TDM~\citep{zhu_2018}, JTM~\citep{zhu2019joint}, AttentionXML~\citep{you2019attentionxml}, and BSAT~\citep{zhuo2020learning}, and graph-based methods like SL2G~\citep{tan2020fast} and NANN~\citep{chen_nann_2022}.
Other works approximate higher-rank interactions using the Mixture of Logits framework and a hierarchical strategy with dot product scoring as the first stage~\citep{zhai_2023,ding_2025}.

Recently, there have been efforts toward in-model index learning for retrieval which discretize continuous item representations into sparse codes -- a concept tracing back to classical latent class models in collaborative filtering~\citep{hofmann1999latent,george2005scalable,agarwal2007predictive} and foundational VQ techniques~\citep{linde1980algorithm}.
In particular, Deep Retrieval~\citep{gao_2020} represents items as discrete paths and learns the item-to-path assignments via Expectation-Maximization.
Inspired by the vector quantized-variational autoencoder (VQ-VAE)~\citep{oord_2017,chen_2020,lee2022autoregressive},~\citet{bin_2025} introduces Streaming VQ for in-model index learning; however, its codebooks are updated by EMA rather than end-to-end supervision from the ranking objective, and its interaction modeling remains restricted to linear dot products.
Other VQ works have focused on quantizing semantic item embeddings for generative recommendation: TIGER~\citep{rajput_2023} and PLUM~\citep{he2026plum} use residual quantization (RQ-VAE) whereas OneRec~\citep{deng2025onerec} uses a residual $k$-means method.

\section{Preliminaries: Retrieval Architecture}\label{sec:prelim}
Modern recommendation systems typically employ a two-stage architecture: retrieval and ranking.
The retrieval stage is recall-focused, modeling $p(\mathrm{engagement}, \mathrm{impression})$ over the entire candidate corpus.
The ranking stage is precision-focused, modeling $p(\mathrm{engagement} \mid \mathrm{impression})$ over the retrieved subset, affording much heavier, early-fused user-item interactions.
The main goal of \oneshot\ is to fundamentally upgrade the retrieval paradigm, so here we focus on the retrieval training and serving formulations, also shown in Fig.~\ref{fig:system_comparison}(a).

{\bf Prediction model training.}
Given user-item embedding pairs $(\vu, \vv)$, the model relies on a contrastive sampled softmax (SSM) loss to discriminate positive items from sampled negatives for a particular engagement signal:
\begin{align}\label{eq:ranking-obj}
\gL_\mathrm{rank}\big( s(\vu, \vv)\big) = - \sum_{(\vu, \vv) \in \mathcal{B}} \log \Big(\frac{e^{\beta {s}({\vu, \vv})}}{e^{\beta {s}({\vu, \vv})} + \sum_{\vv^{-} \in \mathcal{B}_{\vu}^{-}}  e^{\beta {s}({\vu, \vv^-})}}\Big),
\end{align}
where $\beta$ is the inverse softmax temperature, and $\mathcal{B}_{\vu}^{-}$ denotes the pool of the sampled negative items $\vv^-$ for user $\vu$.
Here, dot product $s(\vu, \vv) = \innp{\vu, \vv}$ is the predominant interaction form\footnote{We also correct the logits $s$ via $\log{Q}$ in negative sampling, so that Eq.~\eqref{eq:ranking-obj} provides an unbiased estimate of the full softmax loss across the global candidate corpus~\citep{yi_2019, wu_2024, jean_2015}.}.
While our systems handle multi-task objectives, we drop the task-specific subscripts for clarity.

{\bf Index learning.}
Large-scale retrieval systems traditionally learn the index in a separate step, decoupled from model training under $\gL_\mathrm{rank}$.
Items in the candidate corpus are grouped into $N$ clusters $\{\gS_k\}_{k = 1}^N$ with learned centroids $\{\vc_k\}_{k = 1}^N$ based on the spatial proximity of their embeddings $\vv$, e.g.,\ via $k$-means:
\begin{equation}\label{eq:indexing-obj}
    \gL_{\mathrm{index}}\big(\gS_1,\dots,\gS_N; \vv\big) = \sum_{k = 1}^N \sum_{\vv \in \gS_k} \| \vv - \vc_k \|^2,
\end{equation}
where $\vc_k = \frac{1}{|\gS_k|} \sum_{\vv \in \gS_k} \vv$. While the specific index learning objective varies across different indexing methods, $\gL_\mathrm{index}$ remains fundamentally independent of the primary ranking objective $\gL_\mathrm{rank}$.

Online serving then proceeds in two phases:
first, user embeddings $\vu$ are computed per request and scored against the cluster centroids $\{\vc_k\}_{k = 1}^N$ to select the most relevant clusters;
second, items within these selected clusters are scored against $\vu$ using their embeddings $\vv$ to retrieve a high-recall candidate set.
These steps are essentially the $k$-means ANN algorithm.
While extremely fast, $k$-means ANN is prone to the curse of dimensionality \citep{weber_1998_curse} and suffers from both the rank-index objective misalignment due to the offline $k$-means clustering and an expressiveness limit due to dot-product scoring.

\section{\oneshot: In-model Co-Trained Indexing with Neural Scoring}\label{sec:oneshot}
In this section, we introduce the \oneshot\ framework, which unifies the rank and indexing objectives through index-in-ranking learning ({\it c.f.}\ Eqs.~\eqref{eq:ranking-obj}-\eqref{eq:indexing-obj}) as follows:
\begin{align}\label{eq:index-in-ranking-obj}
    \gL_\text{index-in-rank}\big(s(\vu, \vv_c)\big) = \;& \gL_\mathrm{rank}\Big( s\big(\vu, \underbrace{\mC \bm{\pi}_\mathrm{index}(\vv, \mC)}_{\text{code embedding } \vv_c} \big) \Big),
\end{align}
where $\mC = [ \vc_1, \dots, \vc_{N} ]$ is a trainable codebook and $\bm{\pi}_\mathrm{index}(\vv, \mC)$ is an assignment mapping that outputs an $N$-dimensional hard or soft assignment vector.
While the formulation in Eq.~\eqref{eq:index-in-ranking-obj} may appear straightforward, taking this leap from the traditional decoupled framework to a holistic index-in-ranking scheme introduces non-trivial technical challenges.
For instance, a retrieval inverted index requires a one-hot assignment for each item, which leads to a non-differentiable $\bm{\pi}_\mathrm{index}$.
Further, in-model one-hot encoding often suffers from catastrophic index collapse where a majority of items are assigned to a few indices, which is unacceptable in industrial large-scale retrieval.

In the following, we address these technical challenges and build the index-in-rank scheme.
First, we formalize the in-model hierarchical index and our strategies for E2E co-training using engagement signals and efficient serving using beam search. Next, we present the interaction scale-up scheme with neural scoring functions. Finally, we derive our index-balancing technique based on stochastic compositional optimization theory.

\begin{figure}[t]
    \centering
    \includegraphics[width=0.47\textwidth]{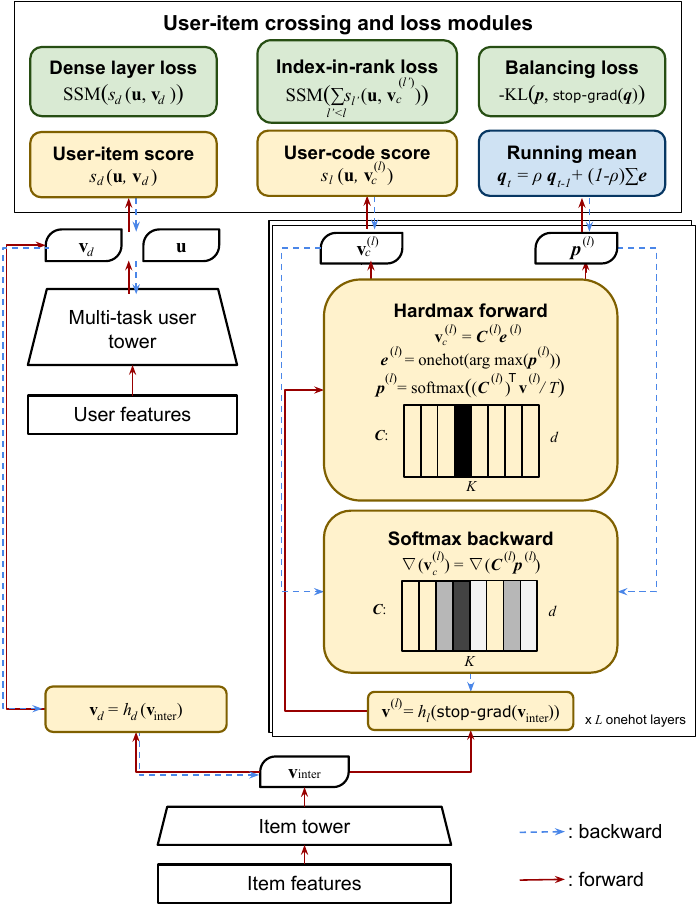}
    \caption{
         \oneshot\ multi-layer one-hot embedding learning architecture. The one-hot encoding $\vv_c^{(\ell)}$ uses hard assignments $\ve^{(\ell)}$ in the forward pass and soft assignment probabilities $\vp^{(\ell)}$ in the backward pass. A Projector component is added to the input of the quantization module with gradient stopping ($\stopgrad$) on the intermediate embedding $\mathbf{\vv_{\text{inter}}}$.
    }
    \label{fig:oneshot_diagram}
\end{figure}
\subsection{In-Model Hierarchical Index}\label{sec:oneshot_indexing}
Extending beyond existing in-model VQ methods, \oneshot\ learns a hierarchy of item one-hot encodings that are fully aligned with recommendation objectives (Figure~\ref{fig:oneshot_diagram}).
These one-hot encodings naturally construct a dynamic index and eliminate the index-ranking misalignment in traditional methods such as $k$-means ANN, HNSW, or NANN.
Furthermore, the one-hot encodings are trained in a boosting fashion with increasing accuracy as the number of layers increases, further enhancing the index efficiency.

To set up index learning, we split the item representation into continuous embeddings $\vv^{(\ell)}$, one for each one-hot layer $\ell$, and a dense embedding $\vv_d$ for full prediction without quantization errors.
Both $\vv^{(\ell)}$ and $\vv_d$ are generated by projecting an intermediate embedding $\vv_{\textrm{inter}}$ from the item tower:
\begin{equation}
\vv^{(\ell)} = h_{\ell}\bigl(\stopgrad(\vv_{\textrm{inter}})\bigr),~\vv_d = h_d(\vv_{\textrm{inter}}).
\label{eq:decouple_item_embeddings}
\end{equation}
The $\stopgrad$ operator prevents the index balancing terms (introduced in Section~\ref{sec:kl_and_objective}) from affecting the learning of $\vv_d$.

\subsubsection{Forward Pass}\label{sec:oneshot_forward_pass}
The hierarchical index contains $L$ layers.
Layer $\ell \in \{1, \dots, L\}$ has a trainable codebook $\mC^{(\ell)} = \bigl[\vc_1^{(\ell)}, \dots, \vc_{N_\ell}^{(\ell)}\bigr] \in \R^{d \times N_\ell}$, whose columns are $N_\ell$ code embeddings of dimension $d$.
Given the continuous item embedding $\vv^{(\ell)}$, the layer computes the code-assignment score vector
\begin{equation}
\vz^{(\ell)} = \bigl(\mC^{(\ell)}\bigr)^\top\vv^{(\ell)},
\label{eq:assignment_score_vector}
\end{equation}
then converts these scores into softmax probabilities and assigns the item to the highest-probability code through one-hot encoding:
\begin{align}
    \vp^{(\ell)}&=\operatorname{softmax}\bigl(\vz^{(\ell)}/T\bigr), \label{eq:soft_assignment}\\
    \ve^{(\ell)}&=\operatorname{onehot}\bigl(\arg\max \vp^{(\ell)}\bigr), \label{eq:hard_assignment}
\end{align}
where $T > 0$ is the temperature.
The selected code embedding is
\begin{equation}
    \vv_c^{(\ell)} = \mC^{(\ell)}\ve^{(\ell)}.
    \label{eq:codebook_mul}
\end{equation}
Each $\vv_c^{(\ell)}$ is used for user-code interaction, and each layer contributes an SSM term to the overall training objective.
Across layers, $\bigl\{\ve^{(\ell)}\bigr\}_{\ell=1}^L$ defines the item's hierarchical code, and this item-to-code mapping is used to construct the inverted index for serving.

In $k$-means terms, $\ve^{(\ell)}$ is an in-model dynamic cluster assignment, and the columns of $\mC^{(\ell)}$ are the trainable cluster centroids.
By jointly optimizing the one-hot item encodings and codebooks with recommendation losses, \oneshot\ unifies representation learning and hierarchical clustering into a single E2E process.
We also introduce a code exploration scheme into this forward pass, but overall, it has minimal effect on the training results (Appendix \ref{sec:gumbel_noise}).
\subsubsection{Backward Pass and Straight-Through Estimator}
Since the one-hot operation in the forward pass is non-differentiable, we employ a Straight-Through Estimator to enable gradient propagation back into the learnable parameters.
Specifically, the backward pass approximates the gradient using the continuous assignment probability $\vp^{(\ell)}$ (Eq.~\ref{eq:soft_assignment}).
We implement this in Eq.~\eqref{eq:codebook_mul} via the stop-gradient operator $\stopgrad$:
\begin{align*}
    \vv_c^{(\ell)} \leftarrow \mC^{(\ell)}\vp^{(\ell)} + \stopgrad\bigl(\mC^{(\ell)}\ve^{(\ell)} - \mC^{(\ell)}\vp^{(\ell)}\bigr).
\end{align*}
This approach preserves the discrete index structure while allowing gradients to reach the learnable parameters.

\subsubsection{Overall Training Objective}
The training loss with \oneshot\ indexing is
\begin{align}\label{eq:loss-oneshot}
    \mathcal{L}_\mathrm{\oneshot}(\vu, \vv) = \;& \sum_{\ell = 1}^L \gL_{\text{index-in-rank}}\Big(\sum_{\ell' \leq \ell} s_{\ell'}\big(\vu, \vv_c^{(\ell')}\big) \Big) \notag \\
    & + \gL_{\mathrm{rank}}\big(s_d(\vu, \vv_d)\big) + \gL_{\mathrm{bal}},
\end{align}
where $s_\ell(\cdot, \cdot)$ and $s_d(\cdot, \cdot)$ denote the scoring functions for computing the logits of user-code and user-dense interactions, and $\gL_{\mathrm{bal}}$ is an index-balancing regularization term over the full candidate corpus to prevent index collapse (detailed in Section~\ref{sec:kl_and_objective}).
The nested summation over $\ell'$ acts as a residual boosting strategy at the logit level, as opposed to the traditional methods of iteratively quantizing residual errors in the same embedding space (VQ-VAE).
This design ensures that each successive layer refines the predictions and improves accuracy over the preceding layers (Section~\ref{sec:exp}).

\subsubsection{Serving Strategies}
\begin{figure*}[ht]
    \centering
    \includegraphics[width=0.75\textwidth]{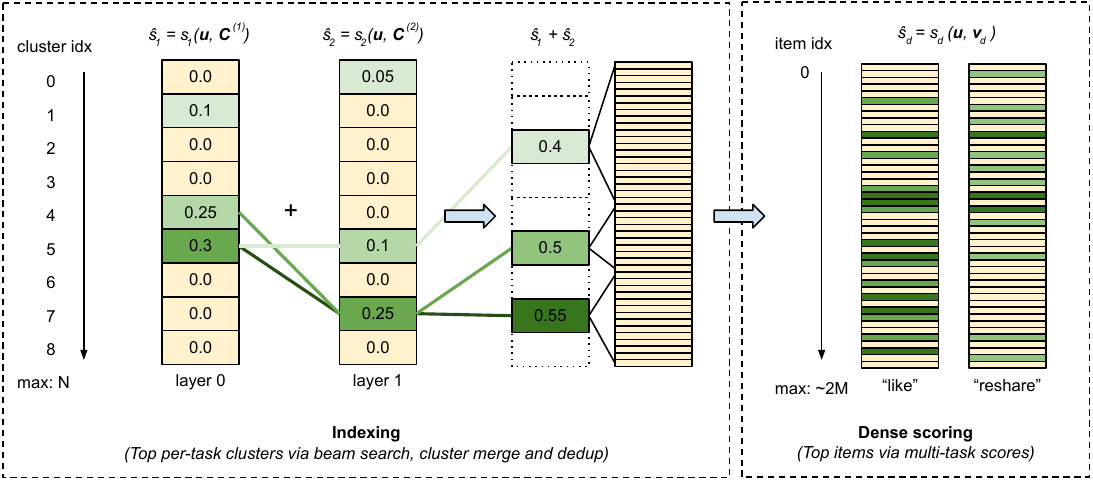}
    \caption{\oneshot\ serving diagram using a 2-layer setting as an example. The order is from left to right, and the colored boxes indicate clusters (large) or items (small). The deeper the green, the higher the scores of clusters or items.}
    \label{fig:oneshot_serving}
\end{figure*}

Serving the multi-layer \oneshot\ model requires a redesign of the traditional ANN pipeline (Appendix Fig.~\ref{fig:oneshot_serving}).
In the index-selection stage, we score the user embedding against each layer's codebook using $s_\ell$, then, consistent with the boosting design, use beam search to find the top code paths for each user based on the summed layer scores along each path\footnote{This path-based selection is similar to Deep Retrieval of ByteDance~\citep{gao_2020}, except that the paths arise naturally as a result of our boosting setup \eqref{eq:loss-oneshot} instead of being trained with Expectation-Maximization. }.
We select the top-$n$ paths for each task and merge their items into a candidate set.
The dense-scoring stage scores every item in the set against the user embedding using $s_{d}(\vu, \vv_d)$ to obtain the final scores.

\subsection{Interaction Scale-Up via Neural Scoring}\label{sec:oneshot_scaleup}
Our framework of in-model indexing with aligned recommendation loss enables architecture scale-up at the indexing stage.
Although prior works have mainly addressed interaction scale-up in ranking models or retrieval models with offline graphs, it has historically been incompatible with in-model indexing frameworks; e.g., see discussions in \citet{bin_2025}.

To achieve interaction scale-up, we use Eq.~\eqref{eq:decouple_item_embeddings} to decouple the code and dense item embeddings, and each index layer is optimized by a recommendation loss on its cumulative code score, while $\vv_d$ is optimized by a separate recommendation loss on $s_d(\vu, \vv_d)$.
We then apply a trainable neural network (NN) on top of the user and item embeddings for the index and dense-scoring stages:
\begin{align*}
s_\ell\big(\vu, \vv_c^{(\ell)}\big) \coloneqq \mathrm{NN}_\ell \big(\vu, \vv_c^{(\ell)}\big), \;
s_d(\vu, \vv_d) \coloneqq \mathrm{NN}_d(\vu, \vv_d),
\end{align*}
where each of $\mathrm{NN}_\ell$ and $\mathrm{NN}_d$ can be any nonlinear neural network that models cross-interactions between user and item embeddings and produces scores for multiple tasks. This makes scaling techniques developed for ranking models, such as cross-attention \citep{rashed_2022_cross_attention} and target-aware mechanisms \citep{li_2025_target_aware}, directly applicable to retrieval.

Additionally, the separation of item one-hot encodings from the dense embedding is necessary to enable interaction scale-up because it frees the one-hot item embeddings from the reconstruction-loss objective used by VQ-VAE and RQ-VAE.
Instead, one-hot encodings produce independent predictions optimized for the final recommendation objectives.
As a result, the one-hot encodings support coarse-grained predictions with a balanced cluster-size distribution, whereas the dense embedding focuses on fine-grained scoring without quantization error.
Compared with Streaming VQ, the dense embedding $\vv_d$ also provides full user-item interaction scores within each cluster rather than relying on a non-personalized item bias term.

During training with minibatch size $B$, each user must be scored against all in-batch items.
Applying a neural network directly to each concatenated user-item pair would require $O(B^2)$ nonlinear evaluations and their intermediate activations, making both computation and memory prohibitive.
We therefore introduce a segment-wise multi-head matrix multiplication to compress the user-item interaction features to a manageable dimension.
Specifically, we reshape $\mathbf{u}$ and $\vv$ into matrices $\mU$ of shape $[d', m]$ and $\mV$ of shape $[d', n]$, and define $s(\mathbf{u}, \vv) = \mathrm{NN}\big(\operatorname{flatten}(\mU^\top \mV)\big), \ s\in\{s_\ell, s_d\}, \mathrm{NN}\in\{\mathrm{NN}_\ell, \mathrm{NN}_d\}$.
In our experiments, we set $d'=256$, $m=12$, and $n=1$.
We show in Section~\ref{sec:offline_result} that scaling up both the width $m$ and the depth $K$ of $\mathrm{NN}_d$ results in increased model performance.

\subsection{Global Index Balancing via Stochastic Compositional Optimization}\label{sec:kl_and_objective}
The two central ingredients of \oneshot\ introduced above, hierarchical in-model indexing and its scale-up via neural scoring, only work if the learned index is selective and well-balanced.
Without explicit balancing, the index can collapse: a small number of clusters or paths absorb most items while many receive almost no assignments.
Indexing collapse compromises codebook expressiveness during training.
More importantly, at serving time, overloaded clusters or paths dominate the index-selection results and significantly hurt serving performance\footnote{Technically, model performance can be recovered if all items from dominating clusters pass through the dense-scoring stage, but the indexing efficiency will be significantly reduced.}.

We note that balancing is already difficult in classical clustering and VQ because the desired target is a global corpus statistic.
Under batch training, a per-batch index-load histogram is only a noisy local estimate rather than a faithful measure of corpus-level balance.
Existing methods therefore rely on direct geometric controls or heuristics.
For $k$-means, $\ell_2$-normalizing points and centroids (spherical $k$-means) can improve cluster balance~\citep{hornik2012spherical}, but this intervention is specific to the Euclidean geometry of the embedding space.
VQ systems use EMA codebook updates paired with dead-code revival or rule-based small-cluster boosting~\citep{oord_2017,dhariwal_2020,bin_2025}.
These are useful engineering tools, but when applied to codebooks co-trained with a recommendation objective, manually overriding assignments or codebook updates can conflict with automatic-differentiation gradients and hurt E2E index learning.

Balancing is even harder in \oneshot\ because item assignments are learned jointly with the recommendation objective rather than determined by a fixed embedding geometry.
In \Cref{sec:oneshot_scaleup}, index scoring is now done by a nonlinear network, so closeness in item embedding space no longer implies closeness in predicted relevance.
Geometry-based balancing can therefore distort assignments preferred by the recommendation model.

To overcome this difficulty, we derive a balancing regularizer $\gL_{\mathrm{bal}}$ in \eqref{eq:loss-oneshot} from a global objective and optimize it jointly with the recommendation losses.
Consider one index layer with $N$ clusters, and let $\ve_{\vj}$ be the hard assignment~\eqref{eq:hard_assignment} of item $j$.
Let $\mathcal{D}_{\mathrm{item}}$ denote the uniform distribution over the deduplicated candidate corpus.
Suppressing the layer index, our optimization target is
\begin{equation}
    \KL\Bigl(
    \mathbb{E}_{\vj \sim \mathcal{D}_{\mathrm{item}}}
    [\ve_{\vj}]
    \;\Big\|\;
    \operatorname{Unif}(N)
    \Bigr).
    \label{eq:layer-kl-composite}
\end{equation}
Here, $\KL(\vq\|\vr)$ denotes the Kullback--Leibler divergence between discrete distributions $\vq$ and $\vr$, and $\operatorname{Unif}(N)$ is the uniform distribution over the $N$ clusters.
In the full loss formulation, \eqref{eq:layer-kl-composite} is applied to every layer.
The expectation is over deduplicated candidate items rather than impressions, because the index should balance the full candidate corpus instead of the skewed user-item interaction data.

The difficulty in optimizing \eqref{eq:layer-kl-composite} is that its gradient depends on the full-corpus average.
Because KL is nonlinear, substituting a per-batch histogram for the inner expectation gives a biased gradient, so minibatch gradients cannot be computed naively.
This nested expectation places \eqref{eq:layer-kl-composite} in the stochastic compositional optimization setting, for which stochastic compositional gradient descent (SCGD) provides an established framework~\citep{wang2017stochastic}.
With careful derivation (given in Appendix~\ref{sec:kl_surrogate_loss_derivation}), we solve this nested objective through a new scalar surrogate loss which makes standard automatic differentiation schemes produce exactly the SCGD parameter gradients. 
This loss can be \emph{directly} added to our existing training objective without the need to change the existing optimizer or implement explicit Jacobians.
The only additional work is updating the corpus-usage state $\hat{\vq}$ and evaluating a vector inner product.
\begin{theorem}[Surrogate realization of soft KL-balancing]
\label{thm:soft-kl-scgd}
For each layer $\ell$, replace the hard assignment $\ve_{\vj}^{(\ell)}$ in \Cref{eq:layer-kl-composite} by the soft assignment probability $\vp_{\vj}^{(\ell)}$, and let $\hat{\vq}^{(\ell)}$ track its corpus average.
Summing these per-layer, or marginal, terms gives the surrogate
\begin{equation}
    \gL_{\mathrm{bal}}%^{\mathrm{marg}}
    =
    \frac{1}{|\gB_{\mathrm{item}}|}
    \sum_{\ell=1}^L \sum_{\vi\in\gB_{\mathrm{item}}}
    \Bigl\langle\vp_\vi^{(\ell)},
    \log \bigl(\stopgrad(\hat{\vq}^{(\ell)})\bigr)\Bigr\rangle,
    \label{eq:kl_surrogate_loss}
\end{equation}
where $\gB_{\mathrm{item}}$ is an item minibatch randomly drawn from deduplicated $\mathcal{D}_{\mathrm{item}}$. 
The automatic differentiation of \eqref{eq:kl_surrogate_loss} gives exactly the SCGD parameter direction for the soft assignment objective. Under appropriate stepsizes, the corresponding SCGD iterates converge almost surely to a stationary point.
\end{theorem}
The above theorem applies to the soft assignment objective, whereas serving-time balance is defined by hard item assignments.
Our implementation therefore retains the same surrogate but updates $\hat{\vq}^{(\ell)}$ from the hard assignments in $\gB_{\mathrm{item}}$.
At a high level, the surrogate loss measures imbalance using the global hard-assignment distribution and applies the balancing pressure by using the soft assignment probabilities as the gradient carrier. 
Appendix~\ref{sec:kl_surrogate_loss_derivation} formulates this use of a hard-assignment tracker as an SCGD update with an STE for the non-differentiable assignments.
For a multi-layer index, the final regularizer $\gL_{\mathrm{bal}}$ combines the per-layer balancing loss \eqref{eq:kl_surrogate_loss} with joint-path balancing, as detailed in Appendix~\ref{sec:joint-path}.

\section{Experiments}\label{sec:exp}
We evaluate three aspects of \oneshot: scaling index expressivity while preserving objective alignment, preventing index collapse, and applicability to other modeling settings.
The experiments comprise the following comparisons and studies:
% Order of importance: 1-layer initial test > multi-layer > balancing > LLM and EID 
\begin{itemize}[wide=0pt, leftmargin=\parindent]
\item We compare the complete \oneshot\ system with the ANN baseline in offline and online evaluations, and study interaction scale-up by varying neural-scoring width and depth;
\item We compare our global index balancing method with existing balancing methods in retrieval and demonstrate its applicability to MoE load balancing;
\item We study prediction refinement across index layers and compare \oneshot-generated item IDs (EID) with offline-computed semantic IDs for generative recommendation.
\end{itemize}
For all offline results reported in this section, we train the retrieval model on 1 day of user impression data (70B rows) and evaluate it on 500K rows sampled from the next day's impression data.
The evaluation uses the full E2E serving pipeline.
\subsection{Offline and Online Performance of \oneshot\ in IG Retrieval}
First, we apply the \oneshot\ in-model indexing and interaction scale-up methods from Section~\ref{sec:oneshot} to our baseline model and evaluate the resulting offline and online performance improvements.
We show that \oneshot\ simultaneously improves the model's asymptotic performance upper bound and indexing efficiency.
\subsubsection{Offline Results}\label{sec:offline_result}
To quantify offline performance improvements, we use impression recall hit rate as an offline evaluation metric.
An impression counts as a hit when the impressed item appears among the model's recommended candidates.
Impression recall hit rate (hereafter, recall) is the number of such hits divided by the total number of impressions.
It quantifies the E2E serving performance of the retrieval model based on users' ground-truth engagements and is a strong indicator of the model's online performance in terms of label click-through rates (CTR) and time-spent.

\begin{figure}[t]
    \centering
    \includegraphics[width=0.47\textwidth]{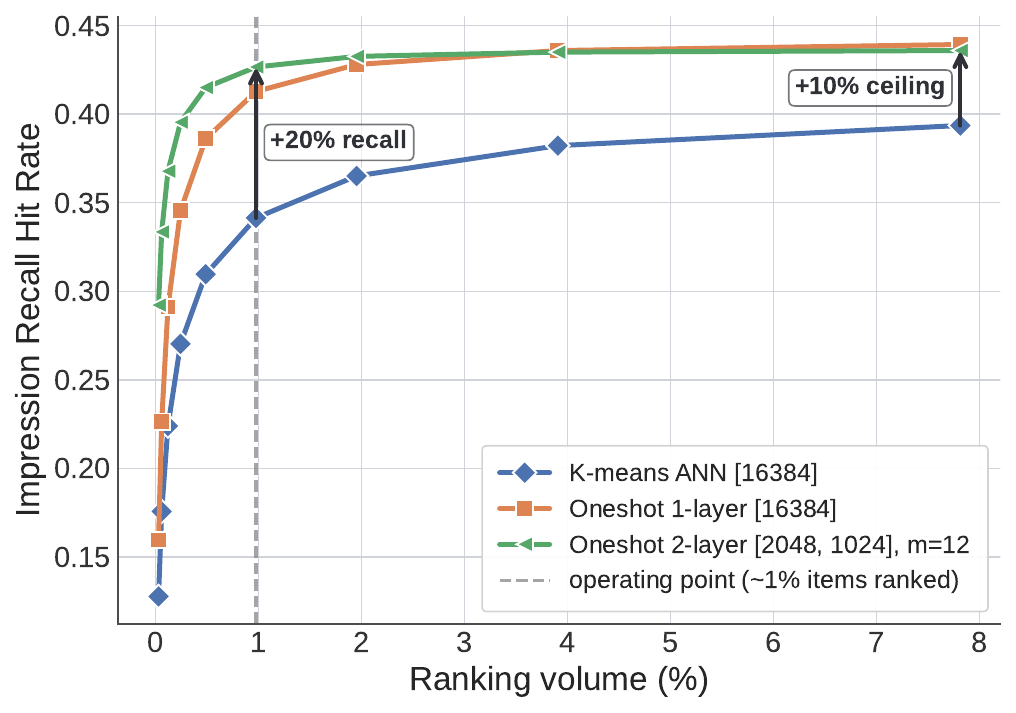}
    \caption{Recall as a function of ranking volume under different \oneshot\ settings. The square brackets in the legend indicate the number of codes in each layer. 
    }
    \vspace{-10pt}
    \label{fig:plot_hit_rate_linear_scale}
\end{figure}

Figure \ref{fig:plot_hit_rate_linear_scale} shows the efficacy of \oneshot\ indexing by comparing the model's impression recall as a function of ranking volume, while keeping the model's output candidate size the same for a fair comparison.
Ranking volume is defined as the percentage of total items ranked at the dense scoring step: a 100\% volume means ranking every item and effectively skipping the indexing step.
At the 1\% operating point, a single-layer \oneshot\ model with 16384 codes has a relative recall improvement of 20\% (0.4128) compared to the baseline with the same number of $k$-means clusters (0.3414).
The \oneshot\ 1-layer setting also yields approximately 10\% larger asymptotic recall when ranking volume increases to more than 8\%, as a result of interaction scale-up.
Moreover, the multi-layer setting with 2 layers $(N_1=2048, N_2=1024)$ achieves the same recall as ANN (selecting 32 clusters) but with 90\% fewer items ranked (with a beam width of 256), achieving a 10x efficiency win.\footnote{Here, ANN has a ranking volume of 32/16384*5=0.976\%, where 5 indicates 5 independent serving tasks, and the 2-layer setting with a beam width of 256 has a ranking volume of 256/1024/2048*5=0.061\%, which is less than 10\% of the ANN's volume.}
The efficiency wins of the multi-layer setting are shown more clearly when the $x$-axis is on a log scale (Figure \ref{fig:plot_hit_rate_log_scale} in Appendix).

\begin{table}[t]
\centering
\caption{Interaction scale-up behavior over increased interaction width $m$ and depth $K$ in $\mathrm{NN}_d$; $K=0$ denotes a simple linear projection from $\mU^\top\mV$ to task logits.}
\label{tab:index_nn_sweep}
\small
\setlength{\tabcolsep}{4pt}
\begin{tabular}{l l c c c}
\toprule
\textbf{Setup} & \textbf{$m$} & \textbf{$K$} & $\mathrm{NN}_d$ \textbf{Params} & \textbf{Recall} \\
\midrule
$k$-means ANN & N/A & N/A & 0 & 0.3701 \\
\midrule
\oneshot & 3 & 0 & 36 & 0.4016 \\
 & 6 & 1 & 105 & 0.4374 \\
 & 12 & 1 & 273 & 0.4519 \\
 & 24 & 2 & 1,425 & 0.4613 \\
\bottomrule
\end{tabular}
\end{table}

We further study the scaling behavior of interaction scale-up by increasing the interaction width $m$ and depth $K$ in the dense-scoring network $\textrm{NN}_d$.
For simplicity, the neural scoring network $\textrm{NN}_d$ uses ReLU activations, and we keep the ranking volume at the 1\% operating point.
Our results show that recall improves gradually as both interaction width and network depth increase (Table~\ref{tab:index_nn_sweep}).
A modest complexity increase from $m=3, K=0$ to $m=24, K=2$ results in a 15\% recall improvement from 0.4016 to 0.4613.
These results suggest that \oneshot\ unlocked the architecture scale-up of the interaction step, and that the scaling behavior of ranking models is now applicable to retrieval.
\subsubsection{Online Results}
We deployed the \oneshot\ 1-layer setting with 16384 codes into global production, achieving multiple top-line wins in users' daily sessions, time-spent, and engagements (Table~\ref{tab:oneshot-online}).
Here, \emph{user sessions} are defined as the number of times a user opens the application, but capped at a daily maximum of 15 to avoid being dominated by power users; a 0.035\% increase at retrieval is a substantial number at our scale.
\emph{Time-spent}, \emph{post impression}, and \emph{like} are defined as the daily total dwell time, number of videos watched, and like button taps; these are operational metrics which support an overall healthy product growth.
\emph{1-Day recency} measures the prevalence of media published within the previous day and indicates the system's responsiveness to new content.

Additionally, we observe a large source rate improvement of 62\%.
\emph{Source rate} is the percentage of impressed items that come from the tested retrieval model, indicating its relative importance in the entire ranking funnel.
Note that the retrieval source with our treatment was already the second largest with an absolute source rate of approximately 20\%, and after the deployment of \oneshot, it became the single largest retrieval source.
The large source rate and engagement improvements indicate that our change is highly aligned with the selection of later-stage ranking models (source rate) as well as users' preferences (engagement).

\begin{table}[t]
\centering
\caption{Online metric improvements of \oneshot\ based on an online A/B test.}
\label{tab:oneshot-online}
\begin{tabular}{l c}
\toprule
\textbf{Metric name} & \textbf{Gain ($\%$)} \\
\midrule
User sessions (capped at 15 daily) & +0.035 \\
Time-spent & +0.136  \\
Post impression & +0.278 \\
Like & +0.251 \\
% Revenue per impression & +0.088 \\
1-Day recency & +5.929 \\
Source rate & +61.58 \\
\bottomrule
\end{tabular}
\end{table}
\subsection{Performance of Global Index Balancing}\label{sec:balancing-exp}
\begin{table*}[ht]
\centering
\caption{Comparison of recall and cluster distribution statistics across different index balancing methods. The cluster size metrics are normalized by the mean cluster size ($\sim$3,430) to illustrate the relative severity of index collapse. The best values in each column are highlighted in \textbf{bold}, and the second-best are \underline{underlined}.}
\label{tab:balancing_results}
\resizebox{\textwidth}{!}{
\begin{tabular}{ll c c c c c}
\toprule
\textbf{Model / Baseline} & \textbf{Balancing Method} & \textbf{Recall} & \textbf{P99/Mean} & \textbf{P999/Mean} & \textbf{Max/Mean} & \textbf{Std/Mean} \\
\midrule
Baseline & None & 0.2420 & 11.52 & 190.16 & 1281.26 & 16.91 \\
$k$-means ANN & Spherical $k$-means ($L_2$) \citep{johnson_2017} & 0.3820 & \underline{2.37} & \underline{3.70} & \textbf{6.66} & \textbf{0.43} \\
VQ-VAE & EMA + dead code revival \citep{dhariwal_2020} & 0.1589 & 23.94 & 42.68 & 66.12 & 4.34 \\
Softmax Correction & Naive softmax correction & 0.2440 & 14.66 & 186.33 & 1395.44 & 16.67 \\
Streaming VQ & EMA + small cluster boost \citep{bin_2025} & \underline{0.4279} & 4.63 & 10.60 & 36.20 & 1.04 \\
\midrule
Ours & \oneshot\ KL-based & \textbf{0.4679} & \textbf{2.26} & \textbf{3.39} & \underline{10.20} & \underline{0.60} \\
\bottomrule
\end{tabular}
}
\end{table*}
We compare our global index balancing method against existing methods when constructing in-model one-hot encodings.
We find that our proposed mechanism dominates the baseline with no balancing treatment and existing heuristics in both E2E recall and cluster distribution statistics (Table \ref{tab:balancing_results}).
\oneshot\ achieves a very uniform cluster distribution, with its p99/mean, p999/mean, max/mean and std/mean metrics comparable to those of the offline $k$-means ANN method, while concurrently achieving a 20\% increase in recall (0.4679 vs 0.3820).
Compared with the baseline run with in-model indexing but no balancing treatment, our method reduces the max/mean cluster-size ratio from the baseline's catastrophic 1281.26 to 10.2.
The baseline also has a dismal recall of 0.2420, suggesting that maintaining a well-balanced cluster distribution is vital for the model's overall serving performance.

Both VQ-VAE with dead-code revival and naive softmax correction (i.e., correcting the current softmax probabilities $\vp$ by subtracting their running mean) severely degrade predictive performance, yielding recalls of 0.1589 and 0.2440, respectively.
Small cluster boosting in Streaming VQ remains competitive with a recall of 0.4279.
However, all these methods still suffer from noticeable tail imbalance, with cluster max/mean growing to 66.12 for VQ-VAE, 1395.44 for naive softmax correction, and 36.2 for streaming VQ; the other balancing statistics are similarly worse than those of our method.

The same balancing problem arises in large Mixture-of-Experts (MoE) language models, where each dense feed-forward layer is replaced by $N$ parallel expert networks and a lightweight router dispatches each token to the top-$k$ experts at that layer~\citep{shazeer_2017, lepikhin_2020, fedus_2022}.
In Appendix~\ref{app:loadbal-losses}, we demonstrate cross-domain applicability of our method, showing that it achieves expert-load distribution statistics on par with existing methods in LLM pretraining.

\subsection{Multi-Layer \oneshot\ Indexing as EID}\label{sec:multi-layer-oneshot}
Finally, we show that prediction accuracy improves progressively across layers and that the multi-layer index recovers most of the predictive ability of the dense embedding.
We also compare the resulting item IDs with traditional semantic IDs in terms of overall retrieval performance to assess the applicability of our IDs to generative recommendation.

\begin{figure}[t]
    \centering
    \includegraphics[width=0.47\textwidth]{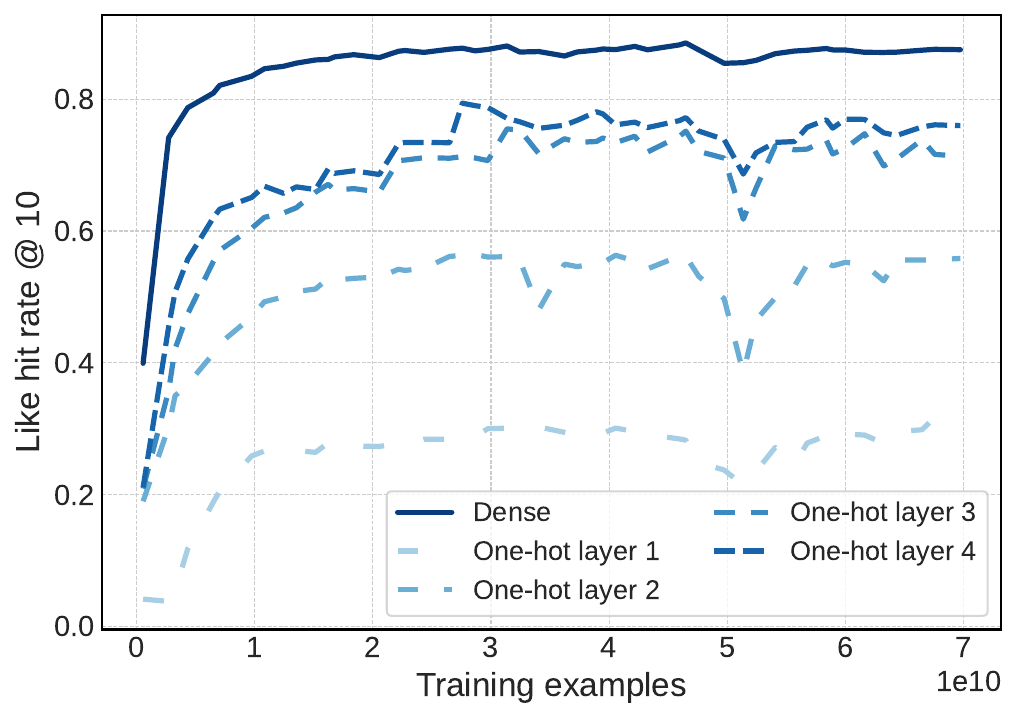}
    \caption{The training trajectory of like hit rate @ 10 for a 4-layer \oneshot\ model with 2048 codes per layer.}
    % \vspace{-10pt}
    \label{fig:4_layer_like_hit_rate}
\end{figure}

Figure \ref{fig:4_layer_like_hit_rate} shows an example of 4-layer one-hot index training with 2048 codes per layer.
We use the train-time metric \emph{like@10 hit rate}\footnote{This is a standard train-time evaluation metric for retrieval models. It is similar to recall but is computed during training for each batch and task; a positive item counts as a "hit" if it ranks among the top 10 over all items in the batch.} to evaluate prediction accuracy as this 4-layer model is impossible to serve due to infrastructure limits. 
Consistent with the boosting design, the performance increases from layer-1 to layer-4.
Importantly, layer-4 recovers roughly 80\% of the dense embedding's performance, indicating that our one-hot encoding can replace the dense embedding given enough inference compute.

We interpret the set of one-hot assignments $\{\ve^{(\ell)}\}_{\ell=1}^L$ as an item ID representation, which we call an Engagement ID (EID).
These in-model EIDs should be more suitable for generic retrieval tasks than semantic IDs because they are co-trained with recommendation objectives and are more expressive, incorporating nearly all item-side information available in the learned dense item embeddings.
EIDs reduce to semantic IDs if we disallow the index-in-ranking objective and restrict item features to SIDs alone.
Table \ref{tab:sid_vs_eid} compares EIDs with state-of-the-art offline-computed SIDs~\citep{zheng_2025, onerec_2025} using the same retrieval model and data.
The model uses two layers with 1024 codes each, matching a two-layer SID with cardinality 1024.
Neural scoring uses width $m=12$ and depth $K=1$.
We highlight the performance gap across three configurations: (A) \textit{SID}, which strictly uses the SID codebook and assignments; (B) \textit{Hybrid}, which retains the codebook but learns the item-to-ID assignment according to the \oneshot\ in-model method; and (C) \textit{\oneshot\ EID}.
\oneshot\ EID significantly outperforms the SID baseline in recall (0.4308 vs. 0.1354) and reduces the max/mean cluster-size ratio from 6254.2 to 300.5, making it more suitable for generative recommendation models.
Interestingly, the \emph{Hybrid} run recovers most of the performance gap with a recall of 0.4230, indicating that learning ID assignments is more effective for retrieval than learning codebooks.

\begin{table}[t]
\centering
\caption{Performance comparison between traditional SIDs and \oneshot\ EIDs.}
\label{tab:sid_vs_eid}
\begin{tabular}{ll c c}
\toprule
\textbf{Config} & \textbf{Codebook, ID} & \textbf{Recall} & \textbf{Max/Mean} \\
\midrule
(A) SID \citep{zheng_2025} & Frozen, Frozen & 0.1354 & 6,254.2 \\
(B) Hybrid & Frozen, Learned & 0.4230 & 402.0 \\
(C) \oneshot\ EID & Learned, Learned & \textbf{0.4308} & \textbf{300.5} \\
\bottomrule
\end{tabular}
\end{table}

% Avoid orphaning the final conclusion sentence on an otherwise empty page.
% Can be removed after we edit the main paper
% \enlargethispage{4\baselineskip}
\section{Conclusion and Discussions}\label{sec:conclusion}
\oneshot\ is an industry-first end-to-end trainable, scalable, and hierarchical indexing framework for large-scale retrieval.
We make four technical contributions.
First, our in-model index is end-to-end trainable and fully aligned with recommendation objectives.
Second, interaction scale-up improves the model's asymptotic performance upper bound.
Third, our loss-based balancing method, derived from compositional optimization theory, yields well-balanced cluster distributions.
Finally, the formulation applies to other ML problems, including semantic ID training for downstream generative recommenders and MoE balancing in LLMs.
We validate these methods by achieving significant offline and online metric gains over our mature, highly optimized industry-scale production baseline after years of active development.

Our framework also opens many possibilities for future work in recommendation.
\oneshot\ engagement IDs substantially outperform offline-generated semantic IDs in retrieval recall.
This result indicates that current models focus on co-engagement information rather than true semantic meaning and that the aligned prediction objectives successfully guide the in-model index toward engagement-based assignments.
A hybrid representation combining semantic and engagement information is therefore likely the best solution for token-based generative recommenders.

\section*{Acknowledgements}
To complete this work, we are extremely grateful to have collaborated with Yiliang Li, Ruofei Shen, Jack Zhao, Xuzhe Zhang, Xiangru Lian, Lei Chen, Liang Wang, Yimin Tan, and Deepak Agarwal. 
% We used AI tools such as Claude Code, Gemini, and ChatGPT to accelerate the coding of this research. 

% \clearpage
% \newpage
\bibliographystyle{assets/plainnat}
\bibliography{ref}

\clearpage
\newpage
\beginappendix

% \section{Supplementary Figures}
\begin{figure}[ht]
    \centering
    \includegraphics[width=0.47\textwidth]{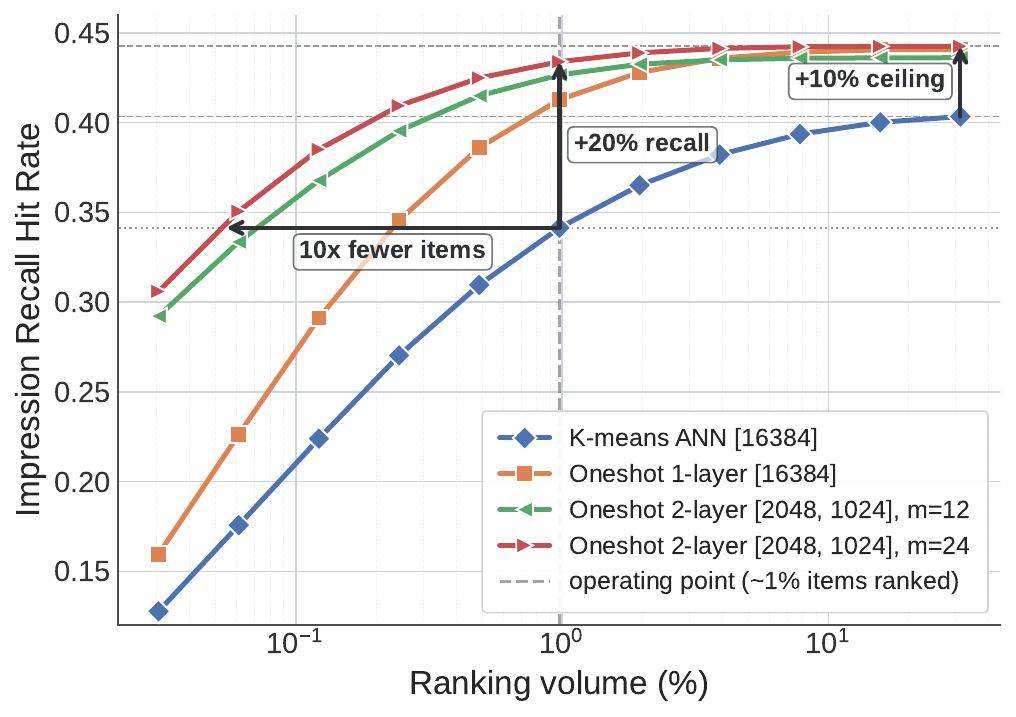}
    \caption{Same as figure \ref{fig:plot_hit_rate_linear_scale} but with log-x scale, and an additional red line which indicates a 2x scale-up of user embedding size $m$. The log scale clearly shows the improved efficiency of multi-layer indexing when the ranking volume is smaller than 1\%.
    }
    \label{fig:plot_hit_rate_log_scale}
\end{figure}

\section{Stochastic Assignments in the Forward Pass}\label{sec:gumbel_noise}
In Section~\ref{sec:oneshot_indexing}, we introduced hard assignment based on $\arg\max$ in the forward pass.
To prevent this hard assignment from stalling exploration during training, we also introduce controlled stochasticity into the soft assignment (Eq.~\ref{eq:soft_assignment}) via the Gumbel-Max trick.
\begin{align*}
\vp^{(\ell)}&=\operatorname{softmax}\bigl(\vz^{(\ell)}/T + \gamma \, \vg^{(\ell)}\bigr),
\end{align*}
where $T$ is the softmax temperature, each $g_{k}^{(\ell)} \sim \mathrm{Gumbel}(0, 1)$ is an independent standard Gumbel random variable, and $\gamma$ modulates the perturbation strength.
The added Gumbel stochasticity exposes the downstream objective to nearby codebook alternatives in \oneshot\ indexing, making the hard assignment behave like sampling from the model's current softmax assignment distribution.
\begin{fact}[Gumbel-Max Trick]
Let $x_1,\dots,x_N$ be unnormalized log-probabilities.
Let $g_1,\dots,g_N$ be independent standard Gumbel random variables.
Then,
$\argmax_{k} \big(x_k + g_k \big) \sim \operatorname{Categorical}\big(\softmax(x_1,\dots,x_N)\big)$.
\end{fact}
\noindent Thus, when $\gamma=1$, the hard assignment is an exact sample from the categorical distribution with probabilities $\operatorname{softmax}\bigl(\vz^{(\ell)}/T\bigr)$.
When $\gamma=0$, the assignment reduces to deterministic argmax selection.

Compared with $\gamma=0$, $\gamma=0.1$ yields a 0.5\% relative recall increase but slightly worsens cluster-size statistics, so its overall effect is small.
This setting is similar to that of \citet{fu_2026}, although the magnitude and empirical effect of Gumbel noise are small in our model.
\section{Convergence and Derivation of the Global Index Balancing Surrogate}\label{sec:kl_surrogate_loss_derivation}
The main text suppresses both the layer index and parameter dependence in \Cref{eq:layer-kl-composite}; here we make them explicit for the derivation.
The stochastic compositional optimization family~\citep{wang2017stochastic} has the abstract form
\begin{equation}\label{eq:scp}
    \min_{\theta} \mathbb{E}_{\xi} \Bigl[ f_{\xi} \Bigl( \mathbb{E}_{\zeta|\xi}[g_{\zeta}(\theta)] \Bigr) \Bigr].
\end{equation}
For layer $\ell$, set $g_{\vj}^{(\ell)}(\theta)=\ve_{\vj}^{(\ell)}(\theta)$ and $f_\ell(\vq)=\KL\Bigl(\vq\Big\|\operatorname{Unif}(N_\ell)\Bigr)$, with $\vj$ sampled uniformly from the deduplicated candidate corpus.
Then \Cref{eq:layer-kl-composite} is the special case of \eqref{eq:scp} with a deterministic outer function $f_\ell$; no outer expectation over user-item pairs is needed.
The balancing objective is added to the SSM losses in \Cref{eq:loss-oneshot} and optimized jointly with them.

SCGD resolves the biased-gradient issue by maintaining a state $y$ for the inner expectation.
Its reference iteration is
\begin{align}
    y_{k+1} &= (1 - \beta_k) y_k + \beta_k g_{\zeta_k}(\theta_k), \label{eq:scgd_tracker}\\
    \theta_{k+1} &= \theta_k - \alpha_k \nabla_\theta g_{\zeta_k}(\theta_k) \nabla_y f_{\xi_k}(y_{k+1}), \label{eq:param_update}
\end{align}
where $\beta_k$ and $\alpha_k$ are the tracking and parameter stepsizes, respectively.
For a vector-valued function $g$, $\nabla_x g(x)$ denotes the transpose of its Jacobian with respect to $x$.
The parameter direction in \eqref{eq:param_update} admits the general scalar surrogate
\begin{equation}
    \widetilde{\gL}_k(\theta)
    =
    \Bigl\langle
        g_{\zeta_k}(\theta),
        \stopgrad\Bigl(
            \nabla_y f_{\xi_k}(y_{k+1})
        \Bigr)
    \Bigr\rangle.
    \label{eq:general-scgd-surrogate}
\end{equation}
Here, $\stopgrad(\cdot)$ is the identity in the forward pass and has zero derivative.
Because the second argument is constant during backpropagation,
\[
    \nabla_\theta\widetilde{\gL}_k(\theta)
    =
    \nabla_\theta g_{\zeta_k}(\theta)
    \nabla_y f_{\xi_k}(y_{k+1}),
\]
exactly the SCGD direction.
Automatic differentiation evaluates this vector--Jacobian product from an ordinary scalar loss, so \eqref{eq:param_update} requires no custom optimizer step or explicit Jacobian.

\paragraph{Proof of \Cref{thm:soft-kl-scgd}.}
With the soft inner map $g_{\vj}^{\prime(\ell)}(\theta)=\vp_{\vj}^{(\ell)}(\theta)$, the objective is an instance of \eqref{eq:scp}.
For $f_\ell(\vq)=\KL\Bigl(\vq\Big\|\operatorname{Unif}(N_\ell)\Bigr)$,
\[
    \nabla_{\vq}f_\ell(\vq)
    = \log \vq + \vone + \log(N_\ell)\vone.
\]
Because $\vp_\vi^{(\ell)}$ lies on the probability simplex, $\nabla_\theta\vp_\vi^{(\ell)}\vone=0$.
The two constant-vector terms therefore vanish from the parameter direction.
Hence,
\begin{align*}
    \nabla_\theta
    \Bigl\langle
        \vp_\vi^{(\ell)},
        \log\Bigl(\stopgrad(\hat{\vq}^{(\ell)})\Bigr)
    \Bigr\rangle
    &= \nabla_\theta\vp_\vi^{(\ell)}\log\hat{\vq}^{(\ell)} \\
    &= \nabla_\theta\vp_\vi^{(\ell)}
       \nabla_{\vq}f_\ell\Bigl(\hat{\vq}^{(\ell)}\Bigr),
\end{align*}
which is exactly the SCGD parameter direction in \eqref{eq:param_update} and the KL specialization of \eqref{eq:general-scgd-surrogate}.
Thus \eqref{eq:kl_surrogate_loss} can be passed to the existing optimizer as an ordinary loss.
The almost-sure stationary-point guarantee follows from Theorem~1 of~\citet{wang2017stochastic} under appropriate stepsizes.

The theorem applies to the soft assignment objective.
The deployed retrieval index instead targets hard item assignments.
For this target, we instantiate \eqref{eq:scgd_tracker} so that $\hat{\vq}^{(\ell)}$ tracks the global hard-assignment distribution $\bar{\vq}^{(\ell)}$.
With a deduplicated item minibatch $\gB_{\mathrm{item}}$ sampled uniformly from the candidate corpus, the tracker update is
\begin{equation}\label{eq:marginal-tracker}
    \hat{\vq}^{(\ell)} \leftarrow \rho \hat{\vq}^{(\ell)} + (1 - \rho)\frac{1}{|\gB_{\mathrm{item}}|}\sum_{\vi \in \gB_{\mathrm{item}}} \ve_{\vi}^{(\ell)}.
\end{equation}
This is the SCGD state update with $\beta_k=1-\rho$, implemented as an EMA of deduplicated hard-assignment usage.
The EMA is the algorithmic state prescribed by SCGD, not a replacement for the inner expectation in \Cref{eq:layer-kl-composite}.

The remaining obstacle is $\nabla_\theta\ve_\vi^{(\ell)}$: because argmax is a step function, its true derivative is zero almost everywhere.
We use the Straight-Through Estimator (STE) as a biased but empirically reliable proxy and replace it with $\nabla_\theta\vp_\vi^{(\ell)}$.
The simplex argument in the proof above again removes every constant multiple of $\vone$, so the same surrogate \eqref{eq:kl_surrogate_loss} exposes the resulting parameter direction to automatic differentiation, while $\stopgrad$ keeps the SCGD state outside backpropagation.

\subsection{Joint Path Balancing}\label{sec:joint-path}
For multi-layer indexing, marginally balancing each layer as in \Cref{eq:kl_surrogate_loss} is necessary but may be insufficient.
Even if every individual codebook layer exhibits uniform usage, the model may still use only a tiny fraction of the available cross-layer paths because assignments across layers are highly correlated.
This can bottleneck serving capacity, as the actual retrieval index is defined by the full code path $(k_1, \dots, k_L)$ rather than any single layer assignment.

To address this, we extend our balancing framework to the joint hard-assignment distribution over the full hierarchical code paths.
Let $N_{\Pi} = \prod_{\ell = 1}^{L} N_\ell$ be the total number of possible paths.
We define the joint one-hot assignment via the Kronecker product ($\otimes$):
\begin{align*}
    \ve_{\vi,\mathrm{joint}} = \ve_\vi^{(1)} \otimes \ve_\vi^{(2)} \otimes \cdots \otimes \ve_\vi^{(L)} \in\{0,1\}^{N_{\Pi}}.
\end{align*}
By applying the same SCGD tracker update to the minibatch joint path frequencies ($\hat{\vq}_{\mathrm{joint}}$) and constructing the corresponding joint soft assignment ($\vp_{\vi,\mathrm{joint}}$), we formulate the joint balancing surrogate identically to the marginal case:
\begin{equation}
    \gL_{\mathrm{bal}}^{\mathrm{joint}} = \frac{1}{|\gB_{\mathrm{item}}|} \sum_{\vi \in \gB_{\mathrm{item}}} \vp_{\vi,\mathrm{joint}}^\top \log \Bigl(\stopgrad(\hat{\vq}_{\mathrm{joint}})\Bigr).
\end{equation}
This joint surrogate has the same SCGD interpretation: gradients flow entirely through the soft assignment probabilities, jointly coupling all layers to push probability mass toward under-utilized full code paths.

\paragraph{Final Balancing Loss.}
The regularizer $\gL_{\mathrm{bal}}$ in \eqref{eq:loss-oneshot} is implemented as a weighted combination of the marginal and joint surrogates:
\begin{equation}\label{eq:joint_loss_coeff}
    \gL_{\mathrm{bal}} = \lambda_{\mathrm{marg}} \gL_{\mathrm{bal}}^{\mathrm{marg}} + \lambda_{\mathrm{joint}} \gL_{\mathrm{bal}}^{\mathrm{joint}}.
\end{equation}
This dual-regularization strategy ensures that the marginal term maintains individual codebook utilization, while the joint term actively regularizes the exponentially larger path space required for scalable multi-layer serving.
The surrogate loss and EMA tracker must use uniformly sampled, deduplicated items from the training or serving candidate corpus; otherwise, duplicate items induce a skewed target distribution.

For multi-layer \oneshot\ indexing, adding the joint path balancing loss in \Cref{eq:joint_loss_coeff} significantly improves cluster balance with an acceptable recall trade-off.
\begin{table}[H]
\centering
\caption{Joint path balancing in a $[512,512]$-cluster model}
\label{tab:joint_kl}
\begin{tabular}{c c c c c}
\toprule
\textbf{ $(\lambda_{\mathrm{marg}}, \lambda_{\mathrm{joint}})$} & \textbf{Recall} & \textbf{Max/Mean}  & \textbf{Std/Mean}\\
\midrule
 (10,0) & \textbf{0.4292} & 165.31 &  3.32 \\
 (5,3) & 0.4268 & \textbf{60.11}  & \textbf{1.88}  \\
\bottomrule
% \textbf{Empty cluster\%} & 26.03\% &  \textbf{9.67}\%
\end{tabular}
\end{table}
\section{Applying \oneshot\ Balancing to MoE Routing}
\label{app:loadbal-losses}
\providecommand{\vepsilon}{\bm{\epsilon}}
Here, we apply the \oneshot\ index-balancing method to MoE routing and compare it with the standard Switch/GShard auxiliary loss~\citep{lepikhin_2020, fedus_2022, du_2022}.
When balancing is downweighted, KL better protects the low-load tail and avoids dead experts; at the usual balancing strength, KL and Switch are effectively tied in language-modeling quality and normalized load statistics.

Consider a sparse MoE layer with $N$ experts and top-$k$ routing.
For a batch of $M$ tokens, the router produces soft probabilities $\vp^\tau$ for token $\tau$ and dispatches it to the selected expert set $\mathcal{T}(\vx^\tau)$.
Let
\[
    \vf_i = \frac{N}{kM}\sum_{\tau=1}^{M}\mathbf{1}[\,i \in \mathcal{T}(\vx^\tau)\,],
    \qquad
    \vp_i = \frac{1}{M}\sum_{\tau=1}^{M}\vp_i^\tau ,
\]
so $\vone^\top \vf=N$ and $\vf_i=1$ under perfect balance.
The factor $k$ accounts for top-$k$ MoE dispatch; retrieval codebook routing is the top-$1$ special case.
The MoE analogue of \Cref{eq:kl_surrogate_loss} is
\begin{equation}\label{eq:moe-kl}
    \gL_{\mathrm{KL}} = \innp{\vp,\ \log \bigl(\stopgrad(\hat{\vf})\bigr)},
\end{equation}
where $\hat{\vf}$ is the detached tracker of the hard dispatch load.
Switch has the same form with a linear transform of the detached load,
\[
    \gL_{\mathrm{Switch}}=\innp{\vp,\stopgrad(\hat{\vf})}.
\]
Thus both are auxiliary losses of the form $\innp{\vp,g\bigl(\stopgrad(\hat{\vf})\bigr)}$.
As in \oneshot\ indexing, hard assignments determine the detached balancing signal, while soft routing probabilities carry gradients to the router.
The training objective combines language-modeling cross-entropy (CE) with a $\lambda_{\mathrm{bal}}$-weighted balancing loss.
We evaluate KL and Switch by pretraining a DeepSeek-V3-style 16B model ($N=64$, $k=6$)~\citep{deepseek_v3_2025} and a Qwen3-style 15B model ($N=128$, $k=8$)~\citep{qwen3_2025} from scratch on the C4 dataset~\citep{raffel_2020}.
Within each architecture, compared runs share data order, initialization, optimizer, and training length and differ only in the balancing configuration.
\paragraph{Reduced coefficient: KL protects valleys.}
To stress-test the losses away from near-uniform loads, we reduce $\lambda_{\mathrm{bal}}$ and sweep the Box--Cox family
\begin{equation}
\begin{aligned}
    g_{r}(\vf) &= \frac{\vf^{r}-1}{r}, \\
    g_{1}(\vf)&=\vf-1\;(\text{Switch}), \qquad
    \lim_{r\to0}g_{r}(\vf)=\log \vf\;(\text{KL}),
\end{aligned}
\label{eq:boxcox}
\end{equation}
where powers are element-wise and $r$ controls the load transformation.
\Cref{tab:moe_boxcox} shows the resulting peak--valley trade-off.
Increasing $r$ from KL through Switch and beyond lowers the peak load (Max/Mean), but it also lowers the valley load (Min/Mean).
KL therefore spends more balancing budget on boosting under-used experts rather than suppressing over-used ones, which is consistent with suppressing dead clusters in retrieval indexing.

\begin{table}[H]
\centering
\caption{Box--Cox sweep with a reduced balancing coefficient at 9.4B tokens. Qwen3 uses $\lambda_{\mathrm{bal}} = 10^{-5}$ and DeepSeek-V3 uses $\lambda_{\mathrm{bal}} = 3\times10^{-6}$. Max/Mean is the peak, Min/Mean the valley, and Dead\% is the fraction of experts with load below $0.1\times$ the mean.}
\label{tab:moe_boxcox}
\resizebox{\columnwidth}{!}{
\begin{tabular}{c ccc ccc}
\toprule
 & \multicolumn{3}{c}{\textbf{Qwen3 15B ($N{=}128$)}} & \multicolumn{3}{c}{\textbf{DeepSeek-V3 16B ($N{=}64$)}} \\
\cmidrule(lr){2-4}\cmidrule(lr){5-7}
$r$ & \textbf{Max/Mean} & \textbf{Min/Mean} & \textbf{Dead\%} & \textbf{Max/Mean} & \textbf{Min/Mean} & \textbf{Dead\%} \\
\midrule
$0$ (KL)     & 2.10 & 0.64 & 0.0 & 2.14 & 0.46 & 0.0 \\
$1$ (Switch) & 1.72 & 0.57 & 0.0 & 2.04 & 0.43 & 1.4 \\
$2$          & 1.61 & 0.53 & 0.0 & 1.86 & 0.36 & 2.2 \\
$3$          & 1.49 & 0.48 & 0.0 & 1.65 & 0.36 & 1.8 \\
\bottomrule
\end{tabular}}
\end{table}

The advantage is clearest on DeepSeek-V3: with the same reduced $\lambda_{\mathrm{bal}}$, KL and Switch have nearly identical language-modeling CE, but Switch leaves dead experts while KL leaves none.
On Qwen3, all settings avoid dead experts, and KL gives the highest valley load.
Larger values of $r$ are useful when the only priority is lowering the peak, but they do so by letting the tail fall further.

\paragraph{Usual coefficient: KL matches Switch.}
\Cref{tab:moe_parity} reports the 56.6B-token results with the usual balancing strength.
In this setting, KL matches Switch on CE and perplexity, and the normalized load statistics are comparable on both architectures.

\begin{table}[ht]
\centering
\caption{MoE load-balancing comparison with the usual balancing strength. CE and perplexity (ppl) measure language-modeling quality (lower is better); Max/Mean and Min/Mean describe the normalized expert-load distribution. DeepSeek-V3 uses $\lambda_{\mathrm{bal}}=10^{-5}$, and Qwen3 uses an annealed $\lambda_{\mathrm{bal}}$ ($10^{-3}\!\to\!5{\times}10^{-4}\!\to\!10^{-4}$).}
\label{tab:moe_parity}
\resizebox{\columnwidth}{!}{
\begin{tabular}{l c c c c}
\toprule
\textbf{Balancing} & \textbf{CE} & \textbf{ppl} & \textbf{Max/Mean} & \textbf{Min/Mean}  \\
\midrule
\multicolumn{5}{l}{\textit{DeepSeek-V3 16B ($N{=}64$, $k{=}6$), 56.6B tokens}} \\
Switch~\citep{fedus_2022}              & 2.372 & 10.72 & 1.46 & 0.67  \\
KL, ours                              & 2.372 & 10.72 & 1.60 & 0.71 \\
\midrule
\multicolumn{5}{l}{\textit{Qwen3 15B ($N{=}128$, $k{=}8$), 56.6B tokens}} \\
Switch~\citep{fedus_2022}  & 2.39 & 11.04 & 1.12 & 0.92 \\
KL, ours    & 2.40 & 11.04 & 1.13 & 0.93 \\
\bottomrule
\end{tabular}}
\end{table}
\paragraph{Why the behavior differs.}
The Box--Cox sweep is useful because it exposes a simple curvature effect.
The marginal sensitivity is $g'_{r}(\vf)=\vf^{r-1}$: KL increases sensitivity on starved experts and reduces it on overloaded experts, while Switch applies equal sensitivity at every load.
Near-perfect balance, however, the losses are nearly identical.
Write $\vp=\softmax(\vz)$ and $\gL_r=\innp{\vp,g_r\bigl(\stopgrad(\vf)\bigr)}$.
Since $\vf$ is detached,
\[
    \nabla_{\vz}\gL_r
    =
    J g_r(\vf),
    \qquad
    J=\operatorname{diag}(\vp)-\vp\vp^\top,
\]
and $J\vone=0$.
Let $\vf=\vone+\vepsilon$ with $\vone^\top\vepsilon=0$.
For each coordinate,
\[
    g_{r}(1+\vepsilon_i)
    =
    \vepsilon_i+\frac{r-1}{2}\vepsilon_i^2
    +\mathcal{O}(|\vepsilon_i|^3),
\]
with the $r\to0$ case following from the Taylor expansion of $\log(1+\vepsilon_i)$.
Therefore, up to constant offsets annihilated by $J\vone=0$,
\begin{equation}\label{eq:boxcox-grad}
    \nabla_{\vz}\gL_{r}
    =
    J\vepsilon+\frac{r-1}{2}J\vepsilon^2+\mathcal{O}(\|\vepsilon\|^3).
\end{equation}
The first-order term is independent of $r$, explaining why KL and Switch match when the load is close to uniform; the second-order term changes the peak--valley preference once $\lambda_{\mathrm{bal}}$ is reduced and the load moves farther from uniform.

\end{document}